\documentclass[prb,a4paper,showpacs,nofootinbib,twocolumn]{revtex4}
\usepackage[dvips]{graphicx}

\newcommand{\ped}[1]{\ensuremath{_{\rm #1}}}
\newcommand{\ape}[1]{\ensuremath{^{\rm #1}}}

\begin{document}

\title{Electrical anisotropy in high-T$\ped{c}$ granular superconductors
in a magnetic field}

\author{D. Daghero}
\author{P. Mazzetti}
\email{mazzetti@polito.it}
\author{A. Stepanescu}
\author{P. Tura}
\affiliation{INFM - Dipartimento di Fisica, Politecnico di Torino
 - C.so Duca degli Abruzzi, 24 - 10129 Torino (TO), Italy}
\author{A. Masoero}
\affiliation{INFM - Dipartimento di Scienze e Tecnologie Avanzate
dell'Universit\`{a} del Piemonte Orientale ``Amedeo Avogadro'' -
Corso Borsalino, 54 - 10131 Alessandria (AL), Italy}
\date{\today}

\begin{abstract}
We propose an analytical model devoted to explain the anisotropy
of the electrical resistance observed below the critical
temperature in granular high-T$\ped{c}$ superconductors submitted
to a magnetic field $\mathbf{H}$. Reported experimental results
obtained on a YBCO sample show that the superconducting transition
occurs in two stages, with a steep drop of the resistance at
$T\ped{c}$ and a subsequent, smoother decrease. In this second
stage, the resistance vs. temperature curve is strongly dependent
not only on the field intensity, but also on the angle between
$\mathbf{H}$ and the macroscopic current density \textbf{j}.

We start from the assumption that the resistance below $T\ped{c}$
is mainly due to the weak links between grains. In the model, weak
links are thought of as flat surface elements separating adjacent
grains.  We calculate the probability for a weak link to undergo
the transition to the resistive state, as a function of the angles
it makes with the external magnetic field $\mathbf{H}$ and the
macroscopic current density $\mathbf{j}$. In doing this, an
important role is given to the strong non-uniformity of the local
magnetic field within the specimen, due to the effect of the
screening supercurrents flowing on the surface of the grains.
Finally, we calculate the electrical resistance of the sample in
the two cases $\mathbf{H}\perp\mathbf{j}$ and $\mathbf{H}\parallel
\mathbf{j}$. The predictions of this simple model turn out to be
in reasonable agreement with reported experimental results
obtained on a YBCO granular specimen.
\end{abstract}

\pacs{74.25.Fy; 74.50.+r}

\maketitle

\section{Introduction}\label{sect:Intro}
If a granular sample of a high-T$\ped{c}$ superconductor (HTSC) is
cooled down to its critical temperature (let us call it
$T\ped{c0}$) in the presence of a magnetic field, its electrical
resistance suddenly falls to a value that can be as low as 30\% of
the normal-state resistance just above $T\ped{c0}$ (see Fig. 5 in
Section~\ref{sect:experimental}). This effect, common to many
cuprates, is due to the superconducting transition of the grains
while the intergrain regions (weak links) remain in the resistive
state \cite{Tinkham}. On further cooling, the sample resistance
gradually decreases, eventually becoming zero at a temperature
$T\ped{c}$ whose value depends on the applied magnetic field. In
this second stage of the superconducting transition, the transport
properties of the specimen are entirely controlled by the weak
links between grains, that can be thought of as S-N-S  Josephson
junctions \cite{Barone} with randomly-distributed critical
energies. According to this approach, when the temperature is
lowered the superconducting wavefunctions of the grains gradually
lock in phase. This gives rise to long-range coherence and finally
to the bulk superconductivity. Of course, the transition of each
junction from the superconducting to the resistive state (or
vice-versa) is controlled by the temperature, the current density
crossing the junction and the local magnetic field. These three
parameters, together with the distribution of the Josephson
critical energies, are thus expected to play a major role in
determining the electrical behaviour of the material. The
transition also depends on wether the magnetic field is applied
 during the cooling of the sample (field cooling, FC) or after cooling
 (zero field cooling, ZFC).  Actually, in the FC case field penetration
and trapping within the grains reduces the transition probability
of the weak links, and consequently the value of the electrical
resistance, by reducing the flux compression factor. This last
quantity will be introduced and discussed in
Section~\ref{sect:model} of this paper.

Finally, in many cases, the electrical properties of granular HTSC
below $T\ped{c0}$ are found to depend on the angle between the
magnetic field $\mathbf{H}$ and the macroscopic current density
$\mathbf{j}$, even if the grains are randomly oriented. This
indicates the existence of an electrical anisotropy of the
material, induced by the application of an external magnetic
field. That also this anisotropy is related to the complex
transition dynamics of the weak link network is demonstrated by
its almost complete absence in high-density polycrystalline
MgB$\ped{2}$ samples, where grains are connected through metallic
contacts and there are no weak links\cite{MgB2}.

In the following, we will propose an explanation for the observed
electrical anisotropy in a zero-field-cooled granular HTSC. The
leading idea of the model is that the screening supercurrents
flowing on the surface of the superconducting grains create a
local ``demagnetising field'' that adds to the external one
creating a strongly non-uniform field distribution in the
integrain regions. We will show that this field distribution gives
rise to a structural anisotropy in the network of resistive weak
links, so that the material behaves as a uniaxially anisotropic
medium for the current transport. By starting from simple
hypotheses, which will be discussed and supported by experimental
evidences, we will develop a simple network model that allows
calculating the electrical resistance of the material as a
function of the magnetic field and of the current density in the
two cases where $\mathbf{H}\perp\mathbf{j}$ and
$\mathbf{H}\parallel\mathbf{j}$. Finally, we will compare the
results of our calculations with an extensive set of experimental
results obtained on a YBCO granular specimen.

\section{The approach to the problem}\label{sect:approach}
Let us focus on a zero-field-cooled (ZFC) granular superconductor.
Suppose to feed it with a current of density $\mathbf{j}$ and to
apply a magnetic field $\mathbf{H}$ such that the bulk
superconductivity is disrupted, but the grains remain in the
Meissner state. As long as the grains exclude the magnetic field,
experiments show that the resistance vs. magnetic field curves are
very nearly reversible. This reversibility shows that it is the
flux pinning within the grains that originates hysteresis, while
flux pinning in the intergrain  regions is negligible. In these
conditions, the electrical resistance is found to depend on
whether the magnetic field $\mathbf{H}$ and the macroscopic
current density $\mathbf{j}$ are parallel or perpendicular to each
other. This anisotropy was already evidenced by some measurements
of critical current \cite{Jones,Glowacki}, resistivity
\cite{Asim,Kilic}, magnetization \cite{Chandran}, power
dissipation \cite{Lopez} and  I-V characteristics
\cite{Blackstead} in different HTSCs. Most of the relevant papers
propose an explanation for the anisotropy based on the
conventional theories of the current-driven vortex motion in the
mixed state. Within this picture, the anisotropy arises from the
fact that the Lorentz force between transport current and vortices
depends on the angle between $\mathbf{H}$ and $\mathbf{j}$.
Actually, this requires that the material behaves as a nearly
homogeneous medium with an effective penetration depth
$\lambda\ped{eff}$, where the magnetic flux penetrates in the form
of vortices as in conventional type-II superconductors. As pointed
out by Ginzburg \emph{et al.} \cite{Ginzburg} this approach is
reasonable as long as $\lambda\ped{eff}$ is much greater than the
average grain size, and fluxons can be taken parallel to the
external field. Even in this case, however, the conventional
theories must be improved -- for example to explain why, in the
``$\mathbf{H}\parallel\mathbf{j}$'' case, the voltage drop across
the specimen is far from being nearly zero. Finally, experimental
results that will be presented in the following section clearly
indicate that the anisotropy gradually vanishes when the
temperature approaches $T\ped{c0}$, and this behavior is not
easily explicable within the picture described so far.

In the following we will propose a completely different
explanation for the electrical anisotropy induced in granular
HTSCs by the magnetic field. We will start by representing  the
material as a set of irregularly shaped grains connected through
thin intergrain regions that behave as weak links. We will think
of these weak links as resistively shunted Josephson junctions
with a perfect ohmic behavior in the normal state -- that is, with
a normal-state resistance independent of both magnetic field and
current, as well as of the angle between them. This assumption
finds again support in the fact that, at temperatures close to
$T\ped{c0}$ -- when almost all the weak links are in the resistive
state -- the anisotropy disappears. This suggests that the
anisotropy is related to the \emph{spatial distribution} of
superconductive and resistive weak links within the material,
rather than to an intrinsic dependence of the intergrain
resistivity on the orientation of $\mathbf{H}$ and $\mathbf{j}$.

The problem now is to understand why the distribution of resistive
weak links should be anisotropic. We argue that this distribution
is mainly determined by the distribution of the local magnetic
field intensity in the intergrain region. As a matter of fact, the
local magnetic field $\mathbf{H\ped{\ell}}$ is given by the
superposition of the external field $\mathbf{H}$ and of the
magnetic field created by the screening supercurrents that flow on
the grains' surfaces. As a result, the local field intensity
$H\ped{\ell}$ in a given weak link can be very different from $H$
-- and can vary very drastically from a weak link to another.
Simple geometrical considerations lead to the conclusion that this
variation is related to the spatial orientation of the weak links
 with respect to
$\mathbf{H}$. The strong non-uniformity of the local field makes
the transition probability of the weak links be anisotropic.
Therefore, the spatial distribution of weak links that undergo the
transition to the resistive state becomes anisotropic as well, and
the material behaves as a uniaxially anisotropic medium for the
current.

The influence of the screening supercurrents of the grains on the
intergrain region was already invoked to explain other interesting
properties of granular superconductors, such as the hysteresis of
the critical current \cite{Evetts,Jackiewicz,Altshuler,Huang} and
the ac magnetization curves \cite{Chandran}. In the original
approach by Evetts and Glowacki \cite{Evetts} it was assumed that
a sufficiently weak magnetic field can be excluded both by grains
and by superconducting ``islands'' bounded by closed paths, called
``rings'', made up of grains connected through weak links with
relatively high critical currents. According to their discussion,
the screening supercurrents flowing along the boundaries of the
superconducting regions create ``flux compression'' in the
surrounding weak links where the field has penetrated, which thus
experience a magnetic field more intense than the applied one. It
is worthwhile to notice that, in that paper, the local magnetic
field outside the superconducting regions was supposed to be
\emph{everywhere} greater than the applied field.

On the contrary, experimental studies of the ac magnetization of
granular HTSCs led Chandran and Chaddah \cite{Chandran} to suggest
that the screening supercurrents flowing on the grain surface give
rise to ``flux compression'' in the weak links laying on planes
parallel to the external field $\mathbf{H}$, and to an almost
complete magnetic shielding  of the weak links perpendicular to
$\mathbf{H}$. In spite of the oversimplification implicit in this
model -- grains are thought to be cubic and arranged in a regular
lattice, as in Fig.1 -- the idea it is based on can be safely
assumed to explain the origin of the magnetic field-induced
anisotropy of the resistance in granular superconductors.

In the present paper, we shall neglect the possible contribution
of superconducting ``rings'' to the non-uniformity of the local
magnetic field in the intergrain regions. Actually, the
experimental results to which we will compare the theoretical
predictions of our model were obtained on a YBCO granular specimen
 with a small critical current density (less than $10^5$A/m$^2$ at
$T=27$~K). A simple calculation shows that the maximum magnetic
field created by this current density flowing on a circular ring
made up of grains and superconducting weak links is definitely
negligible with respect to the magnetic fields considered here,
even if the ring is very small. Notice that also in refs.
\cite{Altshuler,Huang} experimental results were reported,
suggesting that flux trapping or exclusion is mostly due to the
grains, rather than to persistent superconducting loops in the
weak link network.

The model we are going to present in the following arises from a
generalization of the idea by Chandran and Chaddah \cite{Chandran}
to a more realistic case, in which the grains have irregular shape
and the weak links are randomly oriented in space. We will show
that this model quantitatively explains the results of a set of
resistance measurements we carried out on a YBCO granular
specimen.

\section{The model}\label{sect:model}
\subsection{The simplest case}\label{sub:simplest}
\begin{figure}[t]
\vspace{-3mm}
\includegraphics[keepaspectratio,width=\columnwidth]{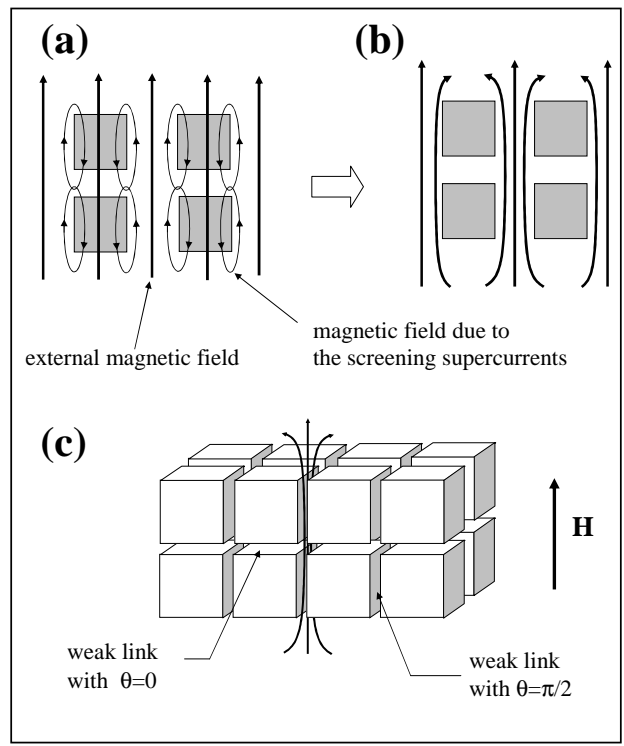}
\vspace{-2.5cm} \caption{In the simplest model for a
polycrystalline HTSC, grains are cubic, arranged in a regular
lattice and separated from one another by intergrain regions that
behave as weak links. (a) If a weak magnetic field (thick arrows)
is applied to the zero-field-cooled material, the screening
supercurrent flowing on the surface of each grain creates a
demagnetizing field (thin loops with arrows). (b) The effective
field in the intergrain region results from the superposition of
the demagnetizing fields of different grains and of the applied
magnetic field. (c) The resulting field pattern is such that the
magnetic field is zero in the weak links with $\theta=0$, and is
greater than the applied one in the weak links with
$\theta=\pi/2$. This field enhancement is usually referred to as
``flux compression''.}\label{fig:cubic}
\end{figure}

To discuss the origin of the field-induced resistance anisotropy
in granular HTSCs, let us start with the analysis of the ideal,
simplified case in which the material is made up of a set of
identical cubic grains arranged in a regular lattice, as shown in
Fig.~\ref{fig:cubic}. Within this simple picture, the weak links
are represented by the flat, square surfaces separating adjacent
grains. Let the magnetic field $\mathbf{H}$ be applied parallel to
one of the grain edges, for example along the vertical direction.
As previously discussed, $\mathbf{H}$ is assumed to be intense
enough to destroy the magnetic screening of the sample as a whole
(due to the supercurrent flowing on the sample surface), but weak
enough not to penetrate into the grains. Let $\theta$ be the angle
between the field $\mathbf{H}$ and the normal $\mathbf{n}$ to a
given weak-link surface. It is clear that, in this simple model,
only the values $\theta=0$ (corresponding to
$\mathbf{n}\parallel\mathbf{H}$) and $\theta=\pi/2$ (corresponding
to $\mathbf{n}\perp\mathbf{H}$) are possible.

The screening supercurrents flowing on the surface of each grain,
which is supposed to be in the Meissner state, create a
``demagnetizing field'' that cancels out the external magnetic
field $\mathbf{H}$ inside the grain. In the surrounding weak
links, the local magnetic field $\mathbf{H\ped{\ell}}$ is given by
the superposition of the demagnetizing fields of adjacent grains
and of the external field $\mathbf{H}$. As shown in Fig.1, the
resulting local magnetic field is much more intense than the
external one (i.e., the flux is ``compressed'') in the weak links
having $\theta=\pi/2$, while it is zero in the weak links having
$\theta=0$.

In the hypothesis that the weak links behave as Josephson
junctions, their transition to the resistive state occurs when the
density of the current crossing the junction is greater than a
critical value $j\ped{c}(T,H)$ which depends on the temperature
and on the local magnetic field.

Therefore, when the current density $\mathbf{j}$ is perpendicular
to the external field $\mathbf{H}$, the condition for the
resistive transition is easily satisfied in the weak links having
$\theta=\pi/2$, which are crossed by the current and submitted to
a strong magnetic field. The resistance of the sample is thus
different from zero.

On the contrary, when $\mathbf{j}$ is parallel to $\mathbf{H}$ the
weak links having $\theta=0$ remain in the superconducting state
even though they are crossed by the current (provided that the
current density is not too large) because of the magnetic
screening of the grains. Therefore, each vertical column of
interconnected grains behaves as a superconducting path for the
current, and the specimen resistance drops to zero.

To summarize, the non-uniformity of the local magnetic field makes
the spatial distribution of resistive weak links depend on the
direction of $\mathbf{j}$ with respect to $\mathbf{H}$. The
macroscopic result is a field-induced anisotropy of the transport
properties of the material (in particular, of the resistivity).

After this simple explanation of the basic mechanism, the
generalization of the model to a more realistic situation, in
which grains have irregular shape and size, has now to be
considered. We shall adopt a statistical point of view and make
some simple assumptions, consistent with the experimental
conditions.

\subsection{The hypotheses}\label{sub:hypotheses}
Let us identify the weak links with flat elements approximating
the surface separating adjacent grains, with random orientation in
space and average area  $\Delta s$.   Let $\mathbf{n}$ be the
unit-length vector normal to their surface, $\theta$ the angle
between $\mathbf{n}$ and the applied magnetic field $\mathbf{H}$,
and $\beta$ the angle between $\mathbf{n}$ and the
\emph{macroscopic} current density $\mathbf{j}$ \footnote{Here and
in the following we shall use $\mathbf{j}$ to indicate the
vectorial average of the current density within the whole
specimen. Instead, the local current density will be indicated by
$\mathbf{j\ped{\ell}}$.}. Let $\mathbf{H\ped{\ell}}$ and
$\mathbf{j\ped{\ell}}$ be the \emph{local} magnetic field and
current density within a weak link. Notice that, as long as the
grains are in the Meissner state, $\mathbf{H\ped{\ell}}$ must be
\emph{parallel} to the weak-link surface -- that is, tangent to
the grain boundary. We will further assume that the weak links
behave as ideal resistively shunted Josephson junctions, with a
perfect ohmic behavior above the transition, and that they have
all the same resistive-state conductance per unit surface, $g$.

\begin{figure}[t]
\includegraphics[keepaspectratio,width=\columnwidth]{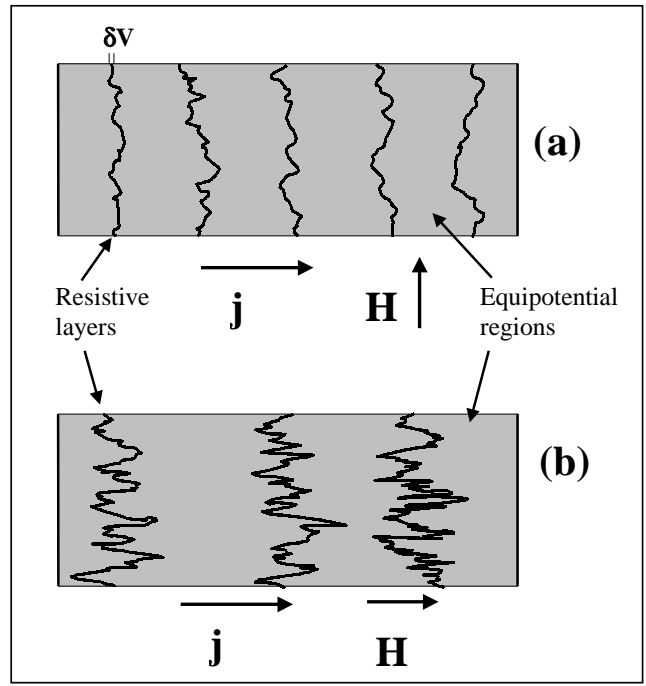}
\vspace{-3cm}\caption{Longitudinal cross section of a
zero-field-cooled cylindrical polycrystalline HTSC some time after
the application of a magnetic field $\mathbf{H}$ intense enough to
make it macroscopically resistive. At the equilibrium, the sample
consists of equipotential superconducting regions (gray),
separated by resistive layers (irregular solid lines) made up of
resistive weak links. The effect of the relative orientation of
$\mathbf{H}$ and $\mathbf{j}$ is made clear by comparing (a) with
(b). It is related to the fact that the probability for a weak
link to undergo the transition to the resistive state is larger
when $\mathbf{H}$ is parallel to its surface, and smaller when
$\mathbf{H}$ is perpendicular to it. }\label{fig:layers}
\end{figure}

In principle, each weak link undergoes the transition from the
superconducting to the resistive state when the local current
density $j\ped{\ell}$ is greater than a critical value
$j\ped{\ell, c}(H\ped{\ell},T)$. However, complex transient
phenomena occurring at the beginning of the conduction process
make it very difficult to determine the actual spatial
distribution of the resistive weak links. These phenomena are due
to correlation effects acting on the local current distribution,
which, for instance, prevent the weak links embedded in a
superconducting region to undergo the resistive transition.
Anyway, when the whole specimen becomes macroscopically resistive,
a stationary situation similar to that sketched in
Fig.~\ref{fig:layers} must be reached. At the equilibrium, the
specimen must be thought of as divided in a set of equipotential
regions separated by resistive layers extended throughout the
specimen cross-section.

The equipotential regions consist of several superconducting
grains and of the weak links between them. Of course, all these
weak links must be in the superconducting state, since any
potential drop within these regions is forbidden. It is worthwhile
to notice that the magnetic flux \emph{may} penetrate within the
weak links embedded in an equipotential region, but the local
current density should remain below its critical value to avoid
the resistive transition.

The resistive layers, instead, are made up of resistive weak
links. Since these layers separate two regions at different
potential, the local current density $\mathbf{j\ped{\ell}}$ that
crosses them must be always \emph{perpendicular} to their surface,
while the local magnetic field $\mathbf{H\ped{\ell}}$ is tangent
to it. Since the weak links are assumed to behave as ideal shunted
Josephson junctions with a perfect ohmic behaviors above the
transition, the electrical resistance of each layer may be
considered as independent of both $\mathbf{H\ped{\ell}}$ and
$\mathbf{j\ped{\ell}}$. As already pointed out, this last
assumption is supported by the resistance vs. temperature curves
reported in Fig.5, measured in a YBCO granular specimen described
in Section~\ref{sect:experimental}.

At a given temperature, the transition of a weak link to the
resistive state is completely determined by $\mathbf{H\ped{\ell}}$
and $\mathbf{j\ped{\ell}}$. As a good approximation, we can say
that the transition occurs when
\begin{equation}\label{eq:jell}
j\ped{\ell} = j\ped{\ell, 0}(T) \frac{H_{0}}{\pi H\ped{\ell}}\;,
\end{equation}
where $j\ped{\ell,0}(T)$ is the value of the local critical
current density in zero magnetic field, and $H_0$ is given by
\[
H_0 =\frac{\phi_0}{4\mu_0\lambda(T)R\ped{g}}\;\;.
\]
Here $\lambda(T)$ is the magnetic penetration depth, $R\ped{g}$ is
the mean radius of the grains, and $\phi_0$ is the flux quantum.
Equation~(\ref{eq:jell}) represents the envelope of the
Fraunhofer-like $I$ vs. $H$ curve of a single Josephson junction
in the presence of a magnetic field parallel to the junction
itself \cite{Barone}. Using the envelope instead of the true
function is usual when a statistical approach is needed -- i.e.,
when a large number of junctions enter into the model. Actually,
as far as the inverse proportionality between $j\ped{\ell}$ and
$H\ped{\ell}$ is concerned, the validity of Eq.~(\ref{eq:jell}) is
supported by experimental results that will be discussed in
section \ref{sect:experimental}.

For any given value $j\ped{\ell}$ of the current density,
Eq.~(\ref{eq:jell}) can be interpreted as a condition on the
intensity of the local magnetic field, $H\ped{\ell}$. The critical
value of $H\ped{\ell}$ giving rise to the transition will be
indicated in the following by
\begin{equation}\label{eq:Hcjell}
H\ped{c}(j\ped{\ell}) = H_0 \frac{j\ped{\ell,0}(T)}{\pi
j\ped{\ell}}\;.
\end{equation}

With reference to the equilibrium situation described in Fig.~2,
the local current density in a resistive weak link is given by
\[
j\ped{\ell}=\mathbf{j}\cdot \mathbf{n}= j \cos{\beta}\;\;.
\]
Thus, Eq.~(\ref{eq:Hcjell}) becomes:
\begin{equation}\label{eq:Hcjell2}
H\ped{c}(j\ped{\ell}) = H_0 \frac{j\ped{\ell,0}(T)}{\pi j}\cdot
\frac{1}{\cos{\beta}} = H\ped{c}(j) \frac{1}{\cos{\beta}}\;.
\end{equation}

We will assume that, for a given value of the local current
density, the critical fields $H\ped{c}(j\ped{\ell})$ of the weak
links follow a gaussian distribution with mean value
$<H\ped{c}(j\ped{\ell})>$ and standard deviation $\sigma\ped{c}$.
In fact, the fluctuation of $H\ped{c}(j\ped{\ell})$ around its
mean value -- within the weak-link ensemble characterized by a
given value of $\mathbf{j\ped{\ell}}$ -- is due to several
uncorrelated causes: the grain orientation mismatch, impurity
segregation at the grain boundaries, non-stoichiometric local
oxygen content, etc. In these cases the assumption that the
fluctuation is gaussian is generally accepted. We further assume
that $\sigma\ped{c}$ is proportional to $<H\ped{c}(j\ped{\ell})>$.
This assumption can be justified by observing that when
$<H\ped{c}(j\ped{\ell})>$ is, for instance, reduced as a
consequence of an increment of $j\ped{\ell}$, also the fluctuation
around its value must change accordingly. Actually, it must be
noticed that $<H\ped{c}(j\ped{\ell})>$ depends on $j\ped{\ell}$
and thus on the spherical angles $\theta$ and $\varphi$, and that
the average is intended to be made over the set of weak links
characterized by given values of these angles.

Equation (\ref{eq:Hcjell2}) indicates that the critical field of
each weak link depends on the angle $\beta$ between its normal
$\mathbf{n}$ and the current density $\mathbf{j}$. However, the
intensity of the local magnetic field in the weak link,
$H\ped{\ell}$, is expected to depend in some way on the angle
$\theta$ between $\mathbf{n}$ and $\mathbf{H}$. That this
dependence should exist is suggested by the simple model of cubic
grains, where the value of the local field was easily determined
for all the allowed values of $\theta$. In the more realistic case
of irregular-shaped grains we are facing here, flux lines meander
through the sample, without violating the requirement of
continuity. It is thus very likely, for instance, that flux lines
are forced to pass through weak links with a very low value of
$\theta$, or that weak links with $\theta$ close to $\pi/2$ are
almost completely screened. Since the value of $H\ped{\ell}$ in a
given weak link characterized by an angle $\theta$ also depends on
the position and on the angle distribution of the nearby weak
links, it is reasonable to assume also in this case a gaussian
distribution of $H\ped{\ell}(\theta)$ around its mean value
$<H\ped{\ell}(\theta)>$ with a variance $\sigma\ped{\ell}$
proportional, as in the previous case, to $<H\ped{\ell}(\theta)>$.
Since $\mathbf{H}\ped{\ell}$ should always be tangent to the grain
surface, we can rather safely assume that the dependence of
$<H\ped{\ell}(\theta)>$ on $\theta$ is expressed by the equation:
\begin{equation}\label{eq:Hltheta}
<H\ped{\ell}(\theta)>= H^{\prime} \sin{\theta}
\end{equation}
analogous to the expression of the local magnetic field intensity
on the surface of a superconducting sphere immersed in a uniform
magnetic field \cite{DeGennes}. In this equation, $H^{\prime}$ is
a constant magnetic field intensity that is related to the
external magnetic field through a ``flux compression factor'' $k$
that takes into account the effect of the flux exclusion by the
grains. Finding an explicit expression for $k$ will be the aim of
the following section.

\subsection{Flux compression factor $k$}\label{sub:kappa}

\begin{figure}[t]
\includegraphics[keepaspectratio,width=\columnwidth]{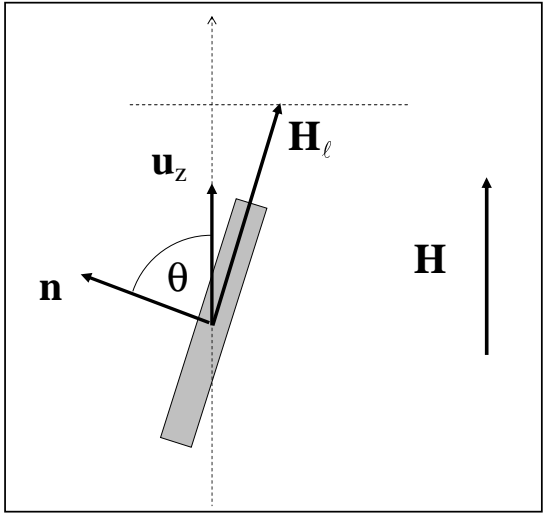}
\vspace{-5.5cm} \caption{Representation of the vectors
characterizing the position of a single weak link. The figure also
shows that the local magnetic field is always parallel to the
weak-link surface. $\theta$ is the angle between the unit-length
vector normal to the surface, $\mathbf{n}$, and the applied
magnetic field $\mathbf{H}$.}\label{fig:3}
\end{figure}

Let's now assume that the magnetic field $\mathbf{H}$ is applied
parallel to the $z$ axis (whose direction is defined by the
unit-length vector $\mathbf{u\ped{z}})$. To determine the value of
$H^{\prime}$, we first write down an expression for the mean value
of the $z$-component of $\mathbf{H\ped{\ell}}$ over the ensemble
of weak links with the same $\theta$. With reference to
Fig.~\ref{fig:3} one finds:
\begin{eqnarray}\label{eq:Hlz}
<H\ped{\ell,z}(\theta)> & = &<H\ped{\ell}(\theta) \sin{\theta} >
\\ &=&
<H\ped{\ell}(\theta)>\sin{\theta}=H^{\prime}(\sin{\theta})^2
\nonumber
\end{eqnarray}

The mean value of the $z$ component over all the weak link
ensemble is now given by a simple solid angle average:
\begin{eqnarray}\label{eq:Hlz2}
<H\ped{\ell,z}> & = & <<H\ped{\ell,z}(\theta)>>_{\theta} \\ & = &
\frac{1}{4\pi} \int H^{\prime} (\sin{\theta})^{2} \mathrm{d}\Omega
= \frac{2}{3}H^{\prime}  \nonumber
\end{eqnarray}

In order to find the compression factor, we express the same
quantity in a different way. In fact, if no superconducting rings
survive, the magnetic field is only excluded by the grains and its
dispersion at the sample edges is expected to be negligible. In
these conditions, the magnetic flux $\Phi$ is very nearly constant
all along the specimen and obviously equal to $H\cdot A_{H}$,
where $H$ is the intensity of the applied magnetic field
$\mathbf{H}$ (which is taken parallel to the $z$ axis) and $A_{H}$
is the specimen cross-section perpendicular to it. On the other
hand, the explicit calculation of the flux gives:
\begin{equation}\label{eq:phi}
\Phi = \int_{A_{H}}\mathbf{H\ped{\ell}}\cdot
\mathbf{u\ped{z}}\cdot \mathrm{d}S =
\int_{A^{\prime}_{H}}H\ped{\ell,z} \cdot \mathrm{d}S =
<H\ped{\ell,z}>A^{\prime}_{H}
\end{equation}

The integral has been restricted to the portion $A^{\prime}_{H}$
of the specimen cross-section in which the magnetic field has
penetrated. Provided that there is no flux penetration into the
grains, $A^{\prime}_{H}$ can be taken as a constant quantity.
Finally, $<H\ped{\ell,z}>$ is obviously the mean value of
$H\ped{\ell,z}$ over $A^{\prime}_{H}$.

Comparing Eq.(\ref{eq:phi}) with the equality $\Phi=H\cdot A_{H}$
gives
\begin{equation}\label{eq:Hlz3}
<H\ped{\ell,z}> = H \cdot \frac{A_{H}}{A^{\prime}_{H}}
\end{equation}
and, by comparing this result with Eq.(\ref{eq:Hlz2}), an
expression for $H^{\prime}$ is finally obtained:

\begin{equation}\label{eq:Hprime}
H^{\prime} = \frac{3}{2} \frac{A_{H}}{A^{\prime}_{H}}H = k H
\hspace{5mm}\text{where}\hspace{5mm} k= \frac{3}{2}
\frac{A_{H}}{A^{\prime}_{H}}
\end{equation}

In conclusion, taking into account Eq.~(\ref{eq:Hltheta}), the
mean value of the local field intensity obtained by averaging over
all the weak links with the same $\theta$ can be written:
\begin{equation}\label{eq:Helltheta}
<H\ped{\ell}(\theta)>= k H \sin{\theta}.
\end{equation}

\subsection{Transition probability for the weak links}\label{sub:transition}
As previously pointed out, we suppose that a weak link with a
given value of $\theta$ undergoes the transition to the resistive
state when the local magnetic field within it,
$H\ped{\ell}(\theta)$, becomes equal to a current-dependent
critical field $H\ped{c}(j\ped{\ell})$. The transition probability
for this weak link is thus:
\begin{equation}\label{eq:Ptrans}
P\ape{tr}(\theta, \beta) = \int_{-\infty}^{+\infty}
f(H\ped{\ell}|\theta)P(H\ped{\ell}\geq
H\ped{c}(j\ped{\ell}))\mathrm{d}H\ped{\ell}
\end{equation}
where $f(H\ped{\ell}|\theta)$ is the distribution of the local
field intensity for a given $\theta$, and $P(H\ped{\ell}\geq
H\ped{c}(j\ped{\ell}))$ is the probability for $H\ped{\ell}$ to be
greater than $H\ped{c}(j\ped{\ell})$. According to previous
assumptions, $f(H\ped{\ell}|\theta)$ can be written:
\begin{equation}\label{eq:f}
f(H\ped{\ell}|\theta) = \frac{1}{\sqrt{2\pi}\sigma\ped{\ell}} \exp
\left[ - \frac{(H\ped{\ell}-k H\sin{\theta})^2}{2
\sigma\ped{\ell}^2} \right]
\end{equation}
where Eq.~(\ref{eq:Helltheta}) has been used to express the mean
value of H$\ped{\ell}$($\theta$). Similarly, the expression for
$P(H\ped{\ell}\geq H\ped{c}(j\ped{\ell}))$ reads:
\begin{eqnarray}\label{eq:P}
P(H\ped{\ell}\geq H\ped{c}(j\ped{\ell}))\! \!&\! =\!\! &\! \!
\int_{0}^{H\ped{\ell}} \!\!\! \frac{1}{\sqrt{2\pi}\sigma\ped{c}}
e^{ - \frac{(H\ped{c}-<H\ped{c}(j\ped{\ell})>)^2}{2
\sigma\ped{c}^2} }\mathrm{d}H\ped{c} \nonumber \\ & = &
\frac{1}{2}\left[ 1+ \mathrm{erf}\left(
\frac{H\ped{\ell}-\frac{<H\ped{c}(j)>}{\cos{\beta}}}{\sqrt{2}\sigma\ped{c}}
\right) \right]
\end{eqnarray}
where Eq.~(\ref{eq:Hcjell2}) has been used to express
$<H\ped{c}(j\ped{\ell})>$ in terms of the angle $\beta$.
\begin{figure}[t]
\includegraphics[keepaspectratio,width=\columnwidth]{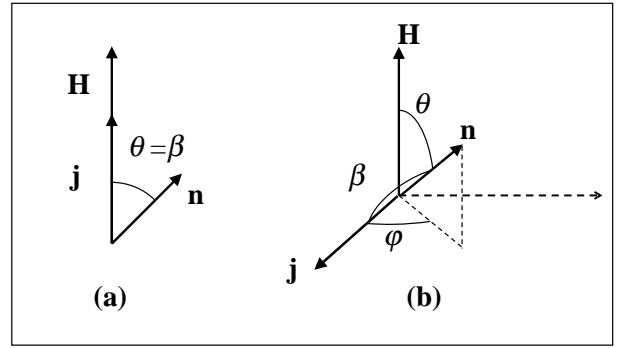}
\vspace{-1.5cm} \caption{Graphic representation of the vectors
$\mathbf{H}$, $\mathbf{j}$ and $\mathbf{n}$ in the two cases where
$\mathbf{H}
\parallel \mathbf{j}$ (a) and $\mathbf{H}
\perp \mathbf{j}$ (b).}\label{fig:4}
\end{figure}
It can be noticed that Eq.~(\ref{eq:Ptrans}) already contains all
the information about the field-induced anisotropy we are dealing
with in the present paper, in the sense that $P\ape{tr}$ will have
a different expression as the current density $\mathbf{j}$ is
parallel or perpendicular to the applied magnetic field
$\mathbf{H}$. As a matter of fact, the relationship between
$\beta$ and $\theta$ is different in the two cases. With reference
to Fig. \ref{fig:4}, it is clear that:
\begin{eqnarray}
\text{if}\,& \mathbf{H} \parallel \mathbf{j} \hspace{1cm}%
& \cos{\beta}=\cos{\theta} \label{eq:parallel}\\
\text{if}\,& \mathbf{H} \perp \mathbf{j} \hspace{1cm} %
& \cos{\beta}= \sin{\theta}\cos{\varphi}\;\;,\label{eq:perp}
\end{eqnarray}
where $\varphi$ is the angle between $\mathbf{j}$ and the plane
containing both $\mathbf{H}$ and $\mathbf{n}$. Taking into account
Eqs.~(\ref{eq:parallel}) and (\ref{eq:perp}), we shall indicate
the transition probability functions for $\mathbf{H} \parallel
\mathbf{j}$ and $\mathbf{H} \perp \mathbf{j}$ as
$P\ped{\parallel}\ape{tr}(\theta)$ and
$P\ped{\perp}\ape{tr}(\theta, \varphi)$, respectively. By using
Eqs.~(\ref{eq:Ptrans}), (\ref{eq:f}) and (\ref{eq:P}), and by
taking into account the expressions for $\cos{\beta}$ given in
Eqs.~(\ref{eq:parallel}) and (\ref{eq:perp}), one obtains
\begin{eqnarray}\label{eq:P_par}
P\ped{\parallel}\ape{tr}(\theta) & = & \int\ped{-\infty}^{+\infty}%
\frac{1}{\sqrt{2 \pi}\sigma\ped{\ell}}%
\exp \left[ - \frac{(H\ped{\ell}-k H %
\sin{\theta})^2}{2\sigma\ped{\ell}^2} \right] \nonumber \\
& \cdot &  \frac{1}{2}\left[ 1+ \mathrm{erf}\left( %
\frac{H\ped{\ell}-\frac{<H\ped{c}(j)>}{\cos{\theta}}}{\sqrt{2}\sigma\ped{c}}%
\right) \right]\mathrm{d}H\ped{\ell}%
\end{eqnarray}
and
\begin{eqnarray}\label{eq:P_perp}
P\ped{\perp}\ape{tr}(\theta,\varphi) & = & \int\ped{-\infty}^{+\infty}%
\frac{1}{\sqrt{2 \pi}\sigma\ped{\ell}}%
\exp \left[ - \frac{(H\ped{\ell}-k H %
\sin{\theta})^2}{2\sigma\ped{\ell}^2} \right] \nonumber \\
& \cdot &  \frac{1}{2}\left[ 1+ \mathrm{erf}\left( %
\frac{H\ped{\ell}-\frac{<H\ped{c}(j)>}{\sin{\theta}\cos{\varphi}}}%
{\sqrt{2}\sigma\ped{c}}%
\right) \right]\mathrm{d}H\ped{\ell}.%
\end{eqnarray}

\subsection{Calculation of the specimen resistance}\label{sub:calculation}
As discussed above, when the specimen resistance is different from
zero, the specimen itself can be described as a series of
equipotential stripes, separated by thin layers made of resistive
weak links. The dependence of the transition probability of the
weak links on their angular position makes these resistive layers
look very different  in the two cases
$\mathbf{H}\parallel\mathbf{j}$ and $\mathbf{H}\perp\mathbf{j}$,
as shown in Fig.~\ref{fig:layers}. Since, in agreement with the
experimental results reported in Fig.~\ref{fig:5}, all the weak
links are assumed to have the same resistive-state conductance per
unit surface $g$, the conductance of each layer is simply
proportional to its area, $S$. The mean value of the area $S$ can
be easily calculated by taking into account that each layer is
made of \emph{resistive} weak links, and that its projection on a
plane perpendicular to the current density \textbf{j} must be
equal to the specimen cross-section area A$_{j}$:
\begin{equation}\label{eq:S}
<S> = \frac{A_{j}}{<|\cos{\beta}|>\ped{res}}
\end{equation}
Here, the subscript ``res'' means that the average is made over
the ensemble of the resistive weak links. Thus:
\begin{equation}\label{eq:cos_beta}
<|\cos{\beta}|>\ped{res} = \frac{\int P\ape{tr}(\theta, \beta)
|\cos{\beta}|\,\mathrm{d}\Omega }{\int P\ape{tr}(\theta, \beta)\,
\mathrm{d}\Omega }\;\;,
\end{equation}
where P$^{tr}$ is given by Eq.(\ref{eq:P_par}) or
Eq.(\ref{eq:P_perp}) according to whether
$\mathbf{H}\parallel\mathbf{j}$ or $\mathbf{H}\perp\mathbf{j}$,
and the integrals are extended to the whole solid angle
$\Omega=4\pi$. The mean value of the resistance of each layer is
thus
\begin{equation}\label{eq:r}
r = \frac{1}{<S> g} = \frac{<|\cos{\beta}|>\ped{res}}{g A_{j}}
\end{equation}

Let $n\ped{L}$ be the number of resistive layers: the resistance
of the specimen is thus given by
\begin{equation}\label{eq:R}
R = n\ped{L} r = n\ped{L} \frac{<|\cos{\beta}|>\ped{res}}{g A_{j}
}.
\end{equation}

To evaluate $n\ped{L}$, we must first calculate the total number
of resistive weak links, $N$, involved in the creation of a given
pattern of resistive layers. $N$ is proportional to the integral
of the transition probability of the weak links, $P\ape{tr}$, over
the whole space. More precisely, if $N\ped{T}$ is the total number
of weak links in the whole specimen, $N$ is given by:
\begin{equation}\label{eq:N}
N = N\ped{T}\int P\ape{tr}(\theta, \beta)\mathrm{d}\Omega.
\end{equation}
On the other hand, the number of weak links that compose a
\emph{single} resistive layer, $n\ped{w}$, is:
\begin{equation}\label{eq:nw}
n\ped{w}= \frac{<S>}{\Delta s} = \frac{A_{j}}{\Delta s \cdot
<|\cos{\beta}|>\ped{res}}
\end{equation}
where $\Delta s$ is the average area of a resistive weak link and
is of the order of the square of the grain radius $R\ped{g}$. Thus
the number of layers results
\begin{equation}\label{eq:nL}
n\ped{L}= \frac{N}{n\ped{w}}= N \frac{\Delta s \cdot
<|\cos{\beta}|>\ped{res}}{A_{j}}
\end{equation}
and the specimen resistance becomes:
\begin{equation}\label{eq:R2}
R = N \frac{\Delta s (<|\cos{\beta}|>\ped{res})^2}{g A_{j}^2}
\end{equation}

It is worthwhile to notice that, in Eq.(\ref{eq:R2}), the specimen
resistance turns out to be proportional to the total number of
resistive weak links. Actually, as is well known from the
percolation theories applied to the superconducting transition
\cite{percolation}, this is not true in the proximity of the
percolative threshold, since below a minimum number of resistive
weak links not a single resistive layer is generated. This implies
that below a given value of $H$ the specimen is in the
superconducting state and obviously Eq.~(\ref{eq:R2}) is not
valid. This point will be taken into account in comparing the
theoretical results with experimental data.

\subsection{Anisotropy of the resistivity}\label{sub:anisotropy}
As previously pointed out, all the information about the
anisotropy of the resistance is already contained in the
transition probability (see Eq.~(\ref{eq:P_par}) and
(\ref{eq:P_perp})). $P\ape{tr}$ enters directly into the
calculation of the specimen resistance through the mean value of
$|\cos{\beta}|$ over the ensemble of resistive weak links. We
shall use $R\ped{\parallel}$ and $R\ped{\perp}$ to indicate the
resistance in the case where $\mathbf{H}\parallel \mathbf{j}$ and
$\mathbf{H}\perp \mathbf{j}$, respectively. According to
Eqs.~(\ref{eq:R2}), (\ref{eq:N}) and (\ref{eq:cos_beta}), and by
using the expressions for $\cos{\beta}$ reported in Eqs.
(\ref{eq:parallel}) and (\ref{eq:perp}), one finds:
\begin{eqnarray}\label{eq:Rparallel}
R\ped{\parallel} = \! \frac{\Delta s}{g A_{j}^2} N\ped{T}\! \!\int
P\ped{\parallel}\ape{tr}(\theta)\mathrm{d}\Omega \cdot \left[
\frac{\int P\ped{\parallel}\ape{tr}(\theta)
|\cos{\theta}|\mathrm{d}\Omega}{\int
P\ped{\parallel}\ape{tr}(\theta) \mathrm{d}\Omega} \right]^2
\end{eqnarray}
and
\begin{eqnarray}\label{eq:Rperp}
R\ped{\perp} & = & \frac{\Delta s}{g A_{j}^2}\cdot N\ped{T}\int
P\ped{\perp}\ape{tr}(\theta,\varphi)\mathrm{d}\Omega \\ & \cdot &
\left[ \frac{\int P\ped{\perp}\ape{tr}(\theta,\varphi)
|\sin{\theta}\cos{\varphi}|\mathrm{d}\Omega}{\int
P\ped{\perp}\ape{tr}(\theta,\varphi) \mathrm{d}\Omega}
\right]^2\;\;. \nonumber
\end{eqnarray}

In order to evaluate the anisotropy of the resistance (that is,
its dependence on the respective orientation of $\mathbf{H}$ and
$\mathbf{j}$) we shall define a parameter $\eta$ such that:
\begin{equation}\label{eq:eta}
\eta =
\frac{R\ped{\parallel}}{R\ped{\perp}}=\frac{\rho\ped{\parallel}}{\rho\ped{\perp}}\;\;.
\end{equation}
The last equality holds because the current always flows  in the
same direction with respect to the specimen (and the direction of
the magnetic field is changed instead). On account of
Eqs.(\ref{eq:Rparallel}) and (\ref{eq:Rperp}), $\eta$ can be
written as follows:
\begin{equation}\label{eq:eta2}
\hspace{-2mm}\eta = \frac{\int P\ped{\perp}\ape{tr}(\theta,
\varphi)\mathrm{d}\Omega} {\int
P\ped{\parallel}\ape{tr}(\theta)\mathrm{d}\Omega} \left[
\frac{\int
P\ped{\parallel}\ape{tr}(\theta)\;|\cos{\theta}|\;\mathrm{d}\Omega}{\int
P\ped{\perp}\ape{tr}(\theta,
\varphi)\;|\sin{\theta}\cos{\varphi}|\; \mathrm{d}\Omega}
\right]^2
\end{equation}
As in the previous section, the integration domain is the whole
solid angle $\Omega$. Actually, when the explicit calculations are
carried out, the symmetries of the problem allow restricting the
$\theta$ and $\varphi$ integrals to the range $[0, \pi/2]$. The
parameters that can be used for fitting the experimental results
are all contained into the functions
P$\ped{\parallel}\ape{tr}(\theta)$ and
P$\ped{\perp}\ape{tr}(\theta, \varphi)$ appearing in the above
expression . One of these parameters is $<H\ped{c}(j)>$, which is
the average of the critical fields over the weak link ensemble,
and can be expressed in terms of the critical current intensity
$j$, according to Eq.~(\ref{eq:Hcjell2}). Another parameter is the
flux-compression coefficient $k$, which, according to
Eq.~(\ref{eq:Helltheta}), simply represents a scaling factor of
the applied magnetic field intensity $H$ in the evaluation of
$<H\ped{\ell}(\theta)>$. As already stated, the standard
deviations $\sigma_{c}$ and $\sigma_{\ell}$ can be taken as
proportional respectively to $<H\ped{c}(j_{\ell})>$ and
$<H\ped{\ell}(\theta)>$, and the proportionality constants become
therefore non-dimensional parameters to be used for the fit.
Actually, as we will see later, the choice of their values has
little influence on the theoretical curves, and thus they can be
considered as non-crucial parameters for the fit. As discussed in
Section~\ref{sect:discussion}, the most important parameter in
fitting the experimental data is $<H\ped{c}(j)>$, which represents
the ``strength'' of the weak link ensemble characteristic of the
specimen under consideration. Results of the numerical
calculations are reported and discussed in
Section~\ref{sect:discussion}.

\section{Experimental results}\label{sect:experimental}
In this section we report the results of an extensive set of
resistance measurements we carried out on a granular YBCO
specimen. The sample, of about $1\times1\times10$~mm$^3$, was
obtained by sinterization of high-purity powders of Y$_2$O$_3$,
BaCO$_3$, and CuO in the stoichiometry ratio $1:2:3$. The oxygen
content of the YBCO chains was then modified by a long-time
annealing (30 days) in a controlled oxygen atmosphere at
$T=720$~K. This process had a twofold effect on the critical
parameters of the material. First, it lowered the critical
temperature down to $T\ped{c0}= 65$~K; second, it reduced the
critical current, that was found to be as low as
2.5$\cdot$10$^5$~A/m$^2$ at $T=27$~K and in zero field. While the
first effect can be ascribed to a change in the oxygen doping in
the grains, the second is mainly due to a strong weakening of the
links between grains. This made the transport properties of the
material below $T\ped{c0}$ be mostly controlled by the weak links
in a wide range of magnetic fields.

The resistance measurements were carried out by using the
conventional four-probe technique. The four contacts were obtained
by Ag evaporation at the opposite ends of the sample. Both the
current and voltage leads were made of thin Pt wires fixed to the
Ag contact by using Ag conductive paste. To eliminate the possible
unwanted contributions of thermoelectric voltages, the
current-reversal technique was used. Moreover, to avoid the small
Joule heating of the sample, the current was injected into the
sample only during the time strictly necessary for the
measurement. The magnetic field was applied either parallel or
perpendicular to the current, which was always flowing along the
same direction, i.e. parallel to the longest side of the sample.
\begin{figure}[t]
\includegraphics[keepaspectratio,width=8cm]{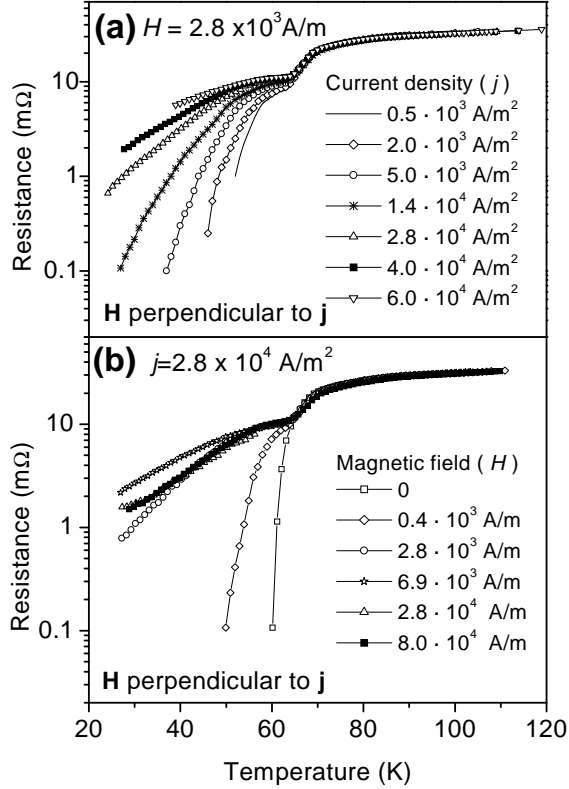}
\caption{Experimental $R$ vs. $T$ curves measured in the YBCO
specimen described in the paper, in the presence of a magnetic
field $\mathbf{H}$ perpendicular to the current density
$\mathbf{j}$. The curves  in (a) were measured with a magnetic
field of intensity $H=2.8\cdot 10^3$~A/m and different values of
the current density. The curves in (b), instead, were obtained
with a current density $j=2.8 \cdot 10^4$~A/m$^2$ and different
values of the applied magnetic field. The drop of the resistance
at the temperature $T\ped{c0}\simeq 65$~K is due to the
superconducting transition of the grains, while the intergrain
regions remain in the resistive state. Notice that the residual
resistance just below $T\ped{c0}$ (that is, the resistance of the
whole weak-link network) is almost independent of both the current
density and the magnetic field.}\label{fig:5}
\end{figure}

The resistance vs. temperature curves reported in Fig.~\ref{fig:5}
were obtained at a fixed value of the magnetic field and different
current densities (a), or at a fixed current and different
magnetic fields (b). In both cases, the magnetic field
$\mathbf{H}$ was perpendicular to the current $\mathbf{j}$.

\begin{figure}[ht]
\includegraphics[keepaspectratio,width=\columnwidth]{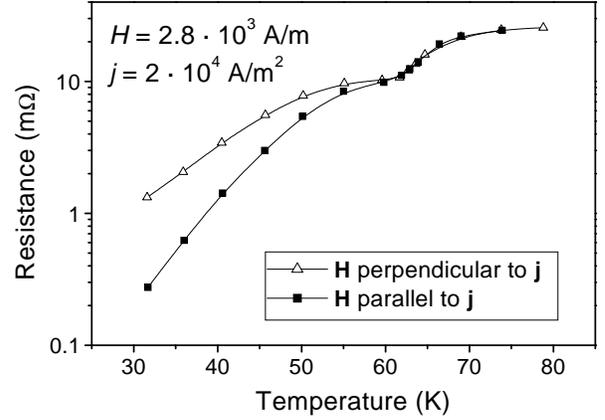}
\caption{Experimental $R$ vs. $T$ curves measured with $j=2\cdot
10^4$~A/m$^2$ and $H=2.8\cdot 10^3$~A/m, in the two cases where
$\mathbf{H}\perp\mathbf{j}$ (open triangles) and
$\mathbf{H}\parallel\mathbf{j}$ (solid squares). Notice that the
residual resistance just below $T\ped{c0}$ is independent of the
orientation of $\mathbf{j}$ with respect to $\mathbf{H}$. As
discussed in the text, this suggests that the anisotropy is not
due to an intrinsic dependence of the intergrain resistivity on
the angle between $\mathbf{H}$ and $\mathbf{j}$.}\label{fig:6}
\end{figure}

The clear step of the curves at the temperature $T\ped{c0}=65$~K
indicates that, at this temperature, the grains become
superconductive while all the weak links remain in the resistive
state. It is clearly seen that the residual resistance just below
$T\ped{c0}$ is practically independent of both the magnetic field
intensity $H$ and the current density $j$. If the temperature is
further lowered, instead, the $R(T)$ curves split depending on the
values of $H$ and $j$.  Fig.~\ref{fig:6} shows the resistance vs.
temperature curves obtained with fixed values of $H$ and $j$ in
the two cases where $\mathbf{H}\parallel \mathbf{j}$ and
$\mathbf{H}\perp \mathbf{j}$. It is clearly seen that the residual
resistance just below $T\ped{c0}$ is as well unaffected by the
respective orientation of these two vectors. Since just below
$T\ped{c0}$ practically all the weak links are still in the
resistive state, this residual resistance can be identified with
that of the \emph{whole} weak-link network. Therefore, these
results indirectly support our hypothesis that the normal-state
resistance of \emph{each} weak link is independent of $\mathbf{H}$
and $\mathbf{j}$ -- and thus on their respective orientation.

\begin{figure}[ht]
\includegraphics[keepaspectratio,width=\columnwidth]{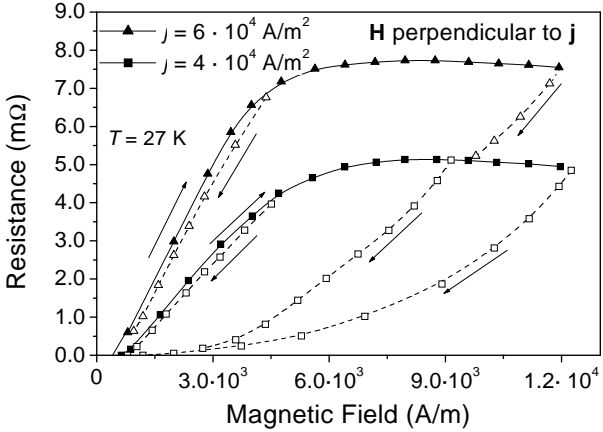}
\caption{Resistance versus magnetic field curves at $T=27$~K for
two different values of the current density: $j=4\cdot
10^4$~A/m$^2$ (squares) and $j=6\cdot 10^4$~A/m$^2$ (triangles).
The magnetic field $\mathbf{H}$ was applied perpendicular to the
current density $\mathbf{j}$. Solid (open) symbols indicate the
resistance measured when the field is increased (decreased), as
indicated by the arrows. The onset of irreversibility at $H
\approx 4.5\cdot 10^3$~A/m is due to flux penetration and flux
trapping inside the superconducting grains. Below this value,
instead, the $R$ vs. $T$ curves are nearly reversible, which
indicates the almost complete absence of flux trapping either by
grains or by superconducting rings. In this regime, the changes in
the resistance due to the magnetic field are attributed to
weak-link transitions. Notice that the curves are in fair
agreement with Eq.~(\ref{eq:Hcjell}), since the same value of the
resistance is obtained for magnetic field values inversely
proportional to the current density $j$.}\label{fig:7}
\end{figure}

In order to study in greater detail the anisotropy of the
resistance highlighted by the curves in Fig.~\ref{fig:6}, and to
compare the predictions of our model with the experimental
results, we measured the resistance of the sample as a function of
the magnetic field intensity, by keeping both the temperature and
the current fixed to a certain value. For example,
Fig.~\ref{fig:7} shows two sets of $R$ vs. $H$ curves obtained
after cooling the sample down to $T=27$~K in zero field, and then
by applying a magnetic field $\mathbf{H}$ perpendicular to
$\mathbf{j}$. The two sets of curves refer to different values of
the current density, $j=4\cdot 10^4$~A/m$^2$ (open and solid
squares) and $j=6\cdot 10^4$~A/m$^2$ (open and solid triangles).
For these values of the current density, self-field effects are
very small and thus we could neglect them.

In Fig.~\ref{fig:7}, solid (open) symbols indicate the resistance
measured while the magnetic field intensity is increased
(decreased). It is clearly seen that the curves are very nearly
reversible up to about $4.5\cdot 10^3$~A/m. In this regime, the
magnetic flux is very likely to be excluded by the grains, and the
variation of the resistance due to the magnetic field can be
ascribed to the transition of weak links from the superconductive
to the resistive state, or vice-versa. Moreover, the curves
reported in Fig.~\ref{fig:7} show that an approximate inverse
proportionality exists between the magnetic field $H$ and the
current density $j$ for a given value of the specimen resistance.
In other words, the same resistance, let's say $R\simeq
2$~m$\Omega$, is obtained with $H\simeq 1.6\cdot 10^3$~A/m and
$j=6\cdot 10^4$A/m$^2$, or with $H\simeq 2.4\cdot 10^3$~A/m and
$j=4\cdot 10^4$A/m$^2$. This is exactly what is expected according
to our model, if Eq. (\ref{eq:Hcjell}) holds true and the sample
is not too close to the percolation threshold.

When the field is increased above $4.5\cdot 10^3$~A/m, the
magnetic flux begins to penetrate into the superconducting grains,
where it remains trapped when the external field is decreased,
thus giving rise to a strong hysteresis. Let us discuss for a
while what happens in this irreversible regime. At the beginning
of the flux penetration into the grains, the increment of the
local magnetic field in the intergrain regions for a given
increase of the external field is smaller than in the reversible
regime. This results in a reduction of the slope of the $R$ versus
$H$ curve. For higher values of the applied field, the flux
penetration into grains gives rise to a decrement of the local
magnetic field, even if the external field is increased, and then
the slope of the curve becomes slightly negative, as clearly shown
in the figure. Within the model developed in the present paper,
this effect can be interpreted as due to a reduction of the flux
compression coefficient $k$, which therefore turns out to depend
on the magnetic field. Actually, we disregarded this dependence
and took $k$ as a constant, which is only true as long as the
magnetic field \emph{does not} penetrate into the grains.
Incidentally, this is one of the reasons why the validity of our
model is restricted to the reversible regime. Let us just point
out here that, if the field is further increased (as we did in
another set of measurements not reported here) the slope becomes
positive again. At about $1.6\cdot 10^5$~A/m, the resistance
becomes practically constant and saturates to a
current-independent value. Notice that the same behaviour can be
observed, close to $T\ped{c0}$, in the curves reported in
Fig.~\ref{fig:5}a.

\begin{figure}[t]
\includegraphics[keepaspectratio,width=\columnwidth]{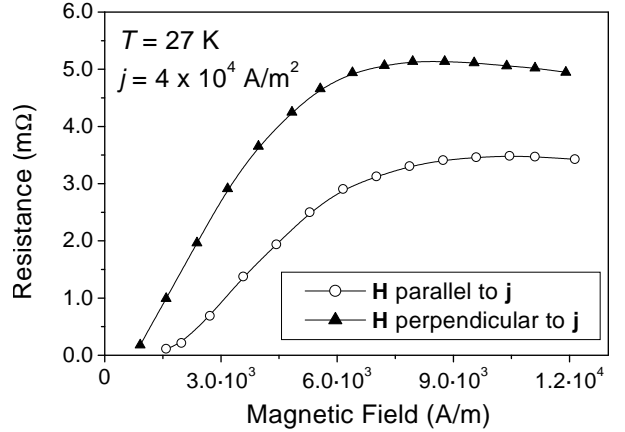}
\caption{Resistance vs. magnetic field curves at the same values
of current density and magnetic field, in the two cases where
$\mathbf{H}\perp \mathbf{j}$ (triangles) and $\mathbf{H}\parallel
\mathbf{j}$ (circles). The anisotropy of the resistance induced by
the magnetic field is clearly seen. The values of $\eta$ reported
in Fig.~\ref{fig:9} are taken from these curves and from a similar
couple of curves measured with a current density $j=6\cdot
10^4$~A/m$^2$.}\label{fig:8}
\end{figure}

\begin{figure}[ht]
\includegraphics[keepaspectratio,width=\columnwidth]{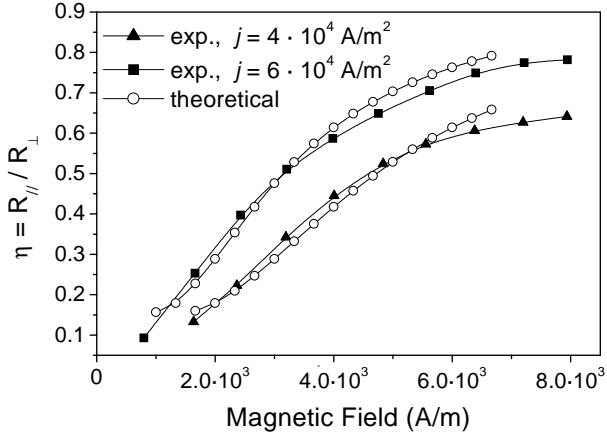}
\caption{Comparison between experimental and theoretical results
concerning the anisotropy factor $\eta$ for two different values
of the macroscopic current density $j$. Solid symbols represent
experimental data, while open circles are theoretical points
calculated by means of Eq.(\ref{eq:eta2}). The theoretical curves
shown here are those that best fit the experimental data, and were
obtained by taking $H\ped{c}^{*}(j)=3.0\cdot 10^3$~A/m  and
$H\ped{c}^{*}(j)=2.0\cdot 10^3$~A/m for $j=4 \cdot 10^4$ and $6
\cdot 10^4$~A/m$^2$, respectively. The ratio between the two
values of $H\ped{c}^{*}(j)$ is clearly the inverse of the ratio
between the current densities, in agreement with
Eq.~(\ref{eq:Hcjell}). The widths of the distributions of critical
fields and local fields were taken as being $\sigma\ped{c}=0.3
\cdot<H\ped{c}(j\ped{\ell})> $ and $\sigma\ped{\ell}=0.7 \cdot \!
\!<H\ped{\ell}(\theta)>$ for both the curves. Notice that the
actual values of $<H\ped{c}(j\ped{\ell})>$ may be an order of
magnitude greater than those of $H\ped{c}^{*}(j)$, owing to the
effect of the flux compression coefficient $k$. }\label{fig:9}
\end{figure}

In Fig.~\ref{fig:8} two $R$ versus $H$ curves are shown, obtained
with the same values of the current density ($4\cdot 10^4$
A/m$^2$) but in the two cases $\mathbf{H}\perp\mathbf{j}$ and
$\mathbf{H}\parallel\mathbf{j}$. By starting from these data sets,
the magnetic field-dependence of the ratio $\eta=R\ped{\parallel}
/ R\ped{\perp}$ can be easily obtained. As a matter of fact, this
dependence is reported in Fig.~\ref{fig:9} (solid triangles)
together with a similar curve obtained with a current density $j=6
\cdot 10^4$~A/m$^2$ (solid squares). In the same figure, the
best-fitting curves calculated by using Eq.~(\ref{eq:eta2}) are
also shown (open circles). The numerical integration of
Eq.~(\ref{eq:eta2}) was performed by means of the computer program
Macsyma V. 2.2, by Macsyma Inc. The values of the best-fit
parameters, reported in the figure caption, will be discussed in
the next section. Let us just point out here that, in spite of the
many simplifications implicit in our model, there is a fair
agreement between theoretical and experimental data.

\section{Discussion and Conclusions}\label{sect:discussion}
The results reported in Fig.~\ref{fig:9} show that there is a
general agreement between the experimental data and the results of
our model. The agreement is particularly good for intermediate
values of the magnetic field intensity. Instead, some deviations
between experimental and theoretical results are evident for low
and high values of the applied field. On account of what was
already pointed out in the previous sections, these deviations are
not surprising. As a matter of fact, our model has been developed
under the assumption that the specimen was neither too close to
the superconductive percolation threshold (that takes place when
the magnetic field is too weak), nor in the condition of flux
penetration within the grains (that occurs when the field is too
intense).

Let us now focus on the best-fitting values of the parameters that
enter in our model. In principle, the adjustable parameters of the
model are: the mean critical field $<H\ped{c}(j)>$, the flux
compression factor $k$, and the standard deviations
$\sigma\ped{\ell}$ and $\sigma\ped{c}$, characterizing the spread
of the local values of $H\ped{\ell}(\theta)$ and $H\ped{c}(j)$.
Since, according to the assumptions made above, $\sigma\ped{\ell}$
and $\sigma\ped{c}$ are taken as being proportional to
$<H\ped{\ell}(\theta)>$ and $<H\ped{c}(j)>$ respectively, it can
be easily shown that $P\ape{tr}\ped{\parallel}(\theta)$ and
$P\ape{tr}\ped{\perp}(\theta,\varphi)$ depend mainly on the
quantity $H\ped{c}^{*}(j) = <H\ped{c}(j)>/ k$ and, to a minor
extent, on the proportionality factors for $\sigma\ped{\ell}$ and
$\sigma\ped{c}$.  $H\ped{c}^{*}(j)$ represents the reduced mean
critical field of the weak links, which depends on the intensity
of the current density, $j$, and on the flux compression factor,
$k$. Thus, according to the present model, comparing the
experimental data with the theoretical results allows the
determination of $H\ped{c}^{*}(j)$ as the main best-fit parameter,
whose value is only little affected by the other two parameters
$\sigma\ped{\ell}$ and $\sigma\ped{c}$. $H\ped{c}(j)$ can thus be
taken as a quantity apt to characterize the ``strength'' of the
weak-links ensemble and it is strongly dependent on the type of
granular superconducting material. It has been observed that
thermal treatments can produce strong changes in this quantity,
leaving practically unaffected the superconducting properties of
the grains, i.e. the critical temperature and the value of the
magnetic field at which the flux begins to penetrate into the
grains \cite{Mazzetti}.

In conclusion, we have presented a simple model which is able to
explain the anisotropy of the resistance shown by granular HTSCs
in the presence of a magnetic field. The model can be applied if
the intensity of the applied magnetic field is such that the bulk
superconductivity is disrupted, but the grains remain in the
superconducting state. This physical requirement is easily
fulfilled in samples where the connections between grains is weak,
as a result, for instance, of thermal treatments and annealing
processes. In these conditions, we have shown that there is no
need of assuming an intrinsic dependence of the intergrain
resistivity on the angle between $\mathbf{H}$ and $\mathbf{j}$, as
instead the models based on the conventional theories of
current-driven flux motion do. Incidentally, the motion of
unpinned vortices is actually the main origin of the resistance in
those materials where the links between grains are so strong that
the flux can penetrate into the grains and even become unpinned
well before the resistive transition of the weak links sets in
\cite{MgB2}.

If the material is made up of weakly linked grains, instead, the
local magnetic field and the local current density in a given weak
link are \emph{always} perpendicular to each other, provided that
the flux penetration into the grains is negligible. In this case
the proposed model, which describes the anisotropy of the
resistivity as being due to the spatially anisotropic distribution
of the resistive weak links seems more appropriate. Actually, the
results of the model are in reasonable agreement with the
experimental data obtained from resistance measurements in a YBCO
granular specimen which satisfies the above mentioned conditions.
Good experimental evidence is also found for the inverse
proportionality between the mean critical field of the weak links
and the current density, which is the basic assumption over which
the proposed model has been developed.

\section*{Acknowledgements}
The authors wish to thank the late prof. P.Manca of the University
of Cagliari for supplying the YBCO sample used for the reported
measurements.

\end{document}